\documentclass[11pt,a4]{article}
\usepackage{amssymb}
\usepackage{amsmath}
\usepackage{amsthm}

\usepackage{hyperref}
\usepackage[dvips]{graphicx,color}

\newcommand{\ket}[1]{\left | #1 \right \rangle}
\newcommand{\bra}[1]{\left \langle #1 \right |}

\def\openone{\leavevmode\hbox{\small1\kern-3.8pt\normalsize1}}

\def\cc{{\cal C}}
\def\cp{{\cal P}}

\def\RR{\mathbb{R}}

\newtheorem{theorem}{Theorem}
\newtheorem{lemma}{Lemma}

\newtheorem{corollary}{Corollary}

\theoremstyle{definition}
\newtheorem{definition}{Definition}

\newcommand{\proj}[1]{\ket{#1}\!\bra{#1}}

\newcommand{\beq}{\begin{equation}}
\newcommand{\eeq}{\end{equation}}
\newcommand{\beqa}{\begin{eqnarray}}
\newcommand{\eeqa}{\end{eqnarray}}

\newcommand{\poly}{{\rm poly}}

\newcommand{\op}[1]{\operatorname{#1}}
\newcommand{\class}[1]{\textup{\textsf{#1}}}
\def\yes{\text{yes}}
\def\no{\text{no}}

\begin{document}

\title{\LARGE\bf
  Matchgate and space-bounded quantum\\
  computations are equivalent}

\author{
  Richard Jozsa$^1$, Barbara Kraus$^2$, Akimasa Miyake$^{2,3}$
  and John Watrous$^4$\\[3mm]
  \small\it
  \small\it $^1$Department of Computer Science, University of
  Bristol,\\[-1mm]
  \small\it Merchant Venturers Building, Bristol BS8 1UB U.K.\\[1mm]
  \small\it $^2$Institute for Theoretical Physics, University of
  Innsbruck,\\[-1mm]
  \small\it Technikerstrasse 25, A-6020 Innsbruck, Austria.\\[1mm]
\small\it
  $^3$Perimeter Institute for Theoretical Physics, \\[-1mm]
  \small\it
  31 Caroline St. N.,Waterloo, Ontario N2L 2Y5, Canada.\\[1mm]
  \small\it $^4$Institute for Quantum Computing and School of Computer
  Science,\\[-1mm]
  \small\it University of Waterloo, 200 University Ave. W., Waterloo,
  Ontario N2L 3G1 Canada.
}

\date{}

\maketitle

\begin{abstract}
  Matchgates are an especially multiflorous class of two-qubit nearest
  neighbour quantum gates, defined by a set of algebraic constraints.
  They occur for example in the theory of perfect matchings of graphs,
  non-interacting fermions, and one-dimensional spin chains.
  We show that the computational power of circuits of matchgates is
  equivalent to that of space-bounded quantum computation with unitary
  gates, with space restricted to being logarithmic in the width of
  the matchgate circuit.
  In particular, for the conventional setting of polynomial-sized
  (logarithmic-space generated) families of matchgate circuits, known to
  be classically simulatable, we characterise their power as
  coinciding with polynomial-time and logarithmic-space bounded
  universal unitary quantum computation.
\end{abstract}

\section{Introduction}\label{intro}

The study of relationships between various kinds of quantum
computational resources is one of the most fundamental issues in
the theory of quantum computation, and one may explore a variety
of possible avenues. On the one hand we may study the power of
quantum computations that use a restricted class of gates or
computational steps (that are generally not fully universal)
e.g.~computations with Clifford gates \cite{cliffgp1,nc} or with
nearest-neighbour matchgates \cite{valclsim,terdiv,jm08}. (The
term ``nearest-neighbour'' here refers to the requirement that the
matchgates act only on consecutive pairs of qubits in the
1-dimensional qubit array of the circuit). Each of these gate
classes is defined by suitable algebraic constraints and in these
cases they lead to computations that turn out to be classically
efficiently simulatable. Despite this classical ceiling on
computing power, these quantum computational processes are still
of considerable interest because, for example, both kinds can
generate complex entangled states and hence illuminate the
sometimes alleged blanket attribution of quantum computational
power to the mere presence of entanglement.

A second approach is to allow unrestricted kinds of (suitably
local, unitary) quantum gates, e.g.~they may be freely chosen from
some finite universal set, and then place restrictions on the
nature and amount of free computational resources that are
available for the computation.
Examples of such resources include time (number of steps), space
(number of qubits), circuit depth (parallel time), deterministic
vs.~non-deterministic computation, etc.
Perhaps the most familiar example of such a restriction is that of
polynomially bounded time, but others that have been established as
conceptually and practically significant in the literature of
{\em classical} complexity theory may also be entertained.

For our present work the notion of (classical or quantum) {\em
space-bounded} computation will be a fundamental ingredient. We
digress here to give a brief intuitive account of it using the
significant and illustrative case of logarithmic space-bounded
computations, called {\em log-space computations} for short. Let
$x = x_1\cdots x_n$ be the binary string input for a computation.
We wish to develop a notion of the computation being carried out
in a restricted space of size only $O(\log n)$. Although this
space is too small to even contain the input, the notion can be
made meaningful and natural as follows. The input is given in an
area of memory called the input tape, that is {\em read only} and
cannot be used for computational processing. (Use of the term
``tape'' here is motivated by a more formal definition based on
the Turing machine model, which is described in the Appendix.)
Computation is then carried out in a separate area of memory
called the work tape, of size $O(\log n)$. As the computation
proceeds, different parts of the input may be read and copied to
the work tape but only a very small, logarithmic length part may
be represented there at any one time. Note, however, that numbers
from 1 to $n$ can be represented with $\log n$ bits, so the
algorithm can at least remember locations in the input and return
to them later if desired. A familiar example of the spirit of
these definitions is our everyday use of the Internet: we may
access desired locations, downloading data for processing on our
computer's hard disk, and remember web addresses, but our hard
disk is far too small to simultaneously contain the whole
Internet.

The above considerations lead to well-defined notions of log-space
computations for both classical and quantum computers, which we
discuss in greater detail in the Appendix. In the classical case
(with deterministic computational steps) any log-space computation
(if it halts) must run in polynomial time \cite{papadim,AB}. In a
similar vein in the quantum case, any log-space quantum
computation may be simulated classically in polynomial time:
$O(\log n)$ qubits correspond to $O(\poly (n))$ dimensions and the
progress of the computation of $N$ steps may then be classically
directly calculated in time $O(N\poly(n))$. Thus, log-space
quantum computation provides another natural class of classically
efficiently simulatable quantum computations.

As mentioned above it is known \cite{valclsim,terdiv,jm08} that
polynomial sized circuits of nearest neighbour (n.n.) matchgates
can be classically efficiently simulated so that their
computational power is at most that of classical poly-time
computation. A main result of the present paper is the precise
identification of their power---we show that it {\em coincides}
with the computational power of quantum log-space computation (in
which the computational steps are unitary operations rather than fully
general trace preserving completely positive maps
\cite{wat1,wat2,wat3}).
Thus we obtain an equivalence between a class of computations
restricted by algebraic constraints (n.n. matchgates) on the one hand,
and a class obtained by limitations on the amount of use of free
computational resources (log-space computation with
general unitary gates) on the other. Actually we prove a more
general result asserting an equivalence (in a precisely defined
sense) between the computational power of general quantum circuits
of size (number of gates) $M$ and width (number of qubit lines)
$m$ on the one hand, and nearest neighbour matchgate circuits of
size $N=O(2^{2m}M)$ and width $n=2^{m+1}$ on the other hand (and
then we can set $m=O(\log n)$ and $M=\poly (n)$ to obtain the
poly-bounded case mentioned above).

Our result is similar in spirit to a theorem of Aaronson and
Gottesman \cite{AG}, who showed that the computational power of
Clifford (or stabilizer) circuits coincides with that of the
classical complexity class known as $\oplus$L (defined in terms of
a certain kind of classical nondeterministic log-space
computation).
In contrast, in our case we have an equivalence between two kinds of
{\em quantum} computations.

In the next section we will recall basic facts about the classical
simulation of nearest-neighbour matchgate circuits, extending
these results for our later purposes and establishing some
notations. In Section \ref{mainresult} we give a precise statement
of our results with reference to an Appendix, which contains the
definitions of space-bounded classical and quantum computation
that we use. In Section \ref{proofs} we give the proof of our main
result, and in Section \ref{ending} some further concluding
remarks.

\section{Quantum matchgate circuits and their classical simulation}
\label{matchsec}

We will follow the notational conventions used in \cite{jm08}.
Let $X,Y,Z$ denote the standard qubit Pauli operators.
A matchgate is defined to be a two-qubit gate $G(A,B)$ of the form (in
the computational basis):
\begin{equation}\label{gab} G(A,B) = \left(
\begin{array}{cccc} p&0&0&q \\ 0&w&x&0 \\ 0&y&z&0 \\ r&0&0&s
\end{array} \right) \hspace{1cm} A = \left( \begin{array}{cc}
p&q \\ r&s \end{array} \right) \hspace{5mm} B= \left(
\begin{array}{cc} w&x \\ y&z \end{array} \right) \end{equation}
where $A$ and $B$ are both in $SU(2)$ or both in $U(2)$ with the
{\em same determinant}. Thus the action of $G(A,B)$  amounts to
$A$ acting in the even parity subspace (spanned by $\ket{00}$ and
$\ket{11}$) and $B$ acting in the odd parity subspace (spanned by
$\ket{01}$ and $\ket{10}$).

For any quantum (matchgate or conventional) circuit $C$, its
{\em size} $N$ is its total number of gates, and its {\em width} $n$
is the total number of qubit lines upon which its gates act.
We will disregard circuits having qubit lines on which no gates
act, so that $N\geq n/2$ for all circuits to be considered.

A fundamental classical simulation result for matchgate circuits
is the following \cite{jm08} (cf. \cite{valclsim,terdiv}).

\begin{theorem}\label{one} Consider any  matchgate circuit of
size $N$ and width $n$, such that:
\begin{enumerate}
\item[(i)]
  the matchgates $G(A,B)$ act on nearest neighbour (n.n.)
  qubit lines only;
\item[(ii)]
  the input state is any computational basis state
  $\ket{x_1\cdots x_n}$;
\item[(iii)]
  the output is a final measurement in the computational basis on
  any single qubit line.
\end{enumerate}
Then the output may be classically efficiently simulated.
More precisely for any $k$ we can classically compute, in $\poly(N)$
time, the expectation value
$\langle Z_k\rangle_{\rm out} = p_0-p_1$ (and hence also $p_0$ and
$p_1$), where $p_0,p_1$ are the outcome probabilities and $Z_k$ is the
Pauli $Z$ operator on the $k$-th line.
\end{theorem}

As in this theorem, two-qubit matchgates in this paper will always
be taken to act only on {\em nearest neighbour} qubit lines so
henceforth the term ``matchgate'' will mean ``nearest neighbour
matchgate''.

For later purposes we will need some details of how this classical
simulation is actually achieved.
We begin by introducing the $2n$ hermitian operators on $n$-qubits
(omitting tensor product symbols $\otimes$ throughout):
\begin{equation}\label{jwr} \begin{array}{cccccc}
c_1=X\,I\cdots I & & c_3= Z\,X\,I\cdots I & \quad\cdots\quad &
c_{2k-1}= Z\cdots Z\,X\,I\cdots I & \quad\cdots
\\
c_2=Y\,I\cdots I & & c_4= Z\,Y\,I\cdots I & \quad\cdots\quad &
\,\,c_{2k}\,\,\,\, = Z\cdots Z\,Y\,I\cdots I & \quad\cdots
\end{array}
\end{equation} where $X$ and $Y$ are in the $k$-th slot for
$c_{2k-1}$ and $c_{2k}$, and $k$ ranges from 1 to $n$.
Thus the operators $c_{2k-1},c_{2k}$ are associated to the $k$-th
qubit line.
It is straightforward to check that these matrices satisfy the
anti-commutation relations
\begin{equation}\label{cliffcomm}
\{ c_j, c_l \} \equiv c_j c_l + c_l c_j = 2
\delta_{j,l} I \hspace{1cm} j,l = 1, \ldots ,2n.
\end{equation}
These relations define a {\em Clifford algebra} $\cc_{2n}$ on $2n$
generators and the operators in eq.~(\ref{jwr}) constitute what is
known as the Jordan-Wigner representation \cite{jwigrep} of the
Clifford algebra.

Next we note the following properties (all proved in \cite{jm08}).
If $U$ is any matchgate acting on lines $(k,k+1)$ then
\begin{equation}\label{mgrot}
  U^\dagger c_j U = \left\{ \begin{array}{ll}
\sum_{l=2k-1}^{2k+2} R[j,l] c_l & \mbox{for
$j=2k-1,2k,2k+1,2k+2$,}
\\ c_j & \mbox{for all other $j$'s,}
\end{array}
  \right.
\end{equation}
where $R\in SO(4,\RR )$ is a special orthogonal matrix and the
$l$-summation extends over the four $c_l$ operators associated to
lines $k$ and $k+1$. Furthermore this association of rotations to
matchgates is surjective, i.e.~every $R\in SO(4,\RR )$ arises from
a matchgate $U$ (unique up to an overall phase). To incorporate
all lines $k$ on an equal footing we can regard $R$ as a rotation
in $2n$ dimensions acting as the identity outside the
4-dimensional span of dimensions $2k-1$, $2k$, $2k+1$ and $2k+2$.

Now let
\[
\ket{\psi_{\rm out}}
= U\ket{\psi_{\rm in}}= U_N \cdots U_1 \ket{x_1\cdots x_n}
\]
represent the action of a circuit of $N$ matchgates on input
$\ket{x_1 \cdots x_n}$.
Then for a final $Z$ measurement on line $k$ we have
\begin{equation}\label{p01a}
  p_0 - p_1
  = \langle Z_k\rangle
  = \bra{\psi_{\rm in}} U^\dagger Z_k U\ket{\psi_{\rm in}}.
\end{equation}
{}From eq.~(\ref{jwr}) it holds that
\[
Z_k= -i c_{2k-1} c_{2k}
\]
and so
\begin{equation}\label{expk}
  U^\dagger Z_k U = -i(U^\dagger c_{2k-1}U)(U^\dagger c_{2k}U).
\end{equation}
Let $R_t$ be the $SO(2n,\RR )$ rotation associated to $U_t$ via
eq.~(\ref{mgrot}) (extended to $2n$ dimensions), and write
$R=R_N\cdots R_1$. Then from eqs.~(\ref{expk},\ref{p01a}) we get
\begin{equation}\label{p01b}
  p_0 - p_1
  = \sum_{j, l = 1}^{2n}
  R[2k-1,j] R[2k,l]
  \bra{\psi_{\rm in}} - i c_{j} c_{l} \ket{\psi_{\rm in}}.
\end{equation}
This finally gives our efficient classical simulation of matchgate
circuits by noting the following features of the formula in
eq.~(\ref{p01b}):
\begin{enumerate}
  \item[(i)]
    $R$ is a product of $N$ matrices of size $2n \times 2n$
    (with $n\leq 2N$) so $R$ can be computed in time $\poly(N)$ by
    sequential matrix multiplication.
  \item[(ii)]
    The $c_j$ operators are all product operators and
    $\ket{\psi_{\rm in}}$ is a product state of $n$ qubits---so
    $\bra{\psi_{\rm in}}c_{j}c_{l}\ket{\psi_{\rm in}}$ can be
    computed as a product of $n$ factors in time $O(n)=O(N)$, for each
    of the $O(n^2)$ choices of $(j,l)$.
  \item[(iii)] The summation in eq.~(\ref{p01b}) has only
    $O(n^2)=O(N^2)$ terms, so in view of (i) and (ii), $p_0-p_1$ can
    be computed in time $\poly(N)$.
    In particular, polynomial-sized matchgate circuits (i.e.~$N=\poly(n)$)
    can be simulated classically in polynomial time.
\end{enumerate}

To develop a relationship of such circuits to space-bounded
quantum computation we begin by giving an alternative expression
of eq.~(\ref{p01b}).
Consider first the case of $k=1$ (i.e.~measurement on the first line)
and $\ket{\psi_{\rm in}} = \ket{0\cdots 0}$.
A direct calculation with the Jordan-Wigner operators gives
\begin{equation}\label{occo}
  \bra{00\cdots 0}-i c_j c_l\ket{00\cdots 0} = \left\{
  \begin{array}{cl}
    1 & \mbox{if $(j,l)=(2k'-1,2k')$} \\
    -1 & \mbox{if $(j,l)=(2k',2k'-1)$} \\
    -i & \mbox{if $j=l$} \\
    0 & \mbox{otherwise.}
  \end{array} \right.
\end{equation}
Because the rows of the matrix $R$ are orthonormal, the sum of terms
in eq.~(\ref{p01b}) with $j = l$ vanishes, and we can write
\begin{equation}\label{Z1ave}
  \langle Z_1\rangle
  = \sum_{j,l = 1}^{2n}
  R[1,j]\,R[2,l]\,S[l,j],
\end{equation}
where we have introduced the matrix of off-diagonal terms from
eq.~(\ref{occo}):
\begin{equation}\label{ess}
  S[l,j] = \left\{
  \begin{array}{cl}
    1 & \mbox{if $(j,l)=(2k'-1,2k')$} \\
    -1 & \mbox{if $(j,l)=(2k',2k'-1)$} \\
    0 & \mbox{otherwise}
  \end{array}
  \right.
\end{equation}
i.e.~$S=\oplus_{i=1}^n \tilde{Y}$ is the block diagonal
anti-symmetric matrix with $n$ copies of
\begin{equation}\label{eqy}
  \tilde{Y} = iY = \left(
  \begin{array}{cc}
    0 & 1 \\ -1 & 0
  \end{array}
  \right)
\end{equation} on the diagonal.

The matrices $R$ and $S$ have size $2n \times 2n$ and real entries,
and using the Dirac notation $\ket{1},\ldots,\ket{2n}$ for an
orthonormal basis in the associated real vector space we can write
eq.~(\ref{Z1ave}) as
\begin{equation}\label{Z1rebit}
  \langle Z_1\rangle =
  \bra{2}RSR^{-1} \ket{1}
\end{equation}
(where we have used the fact that the transpose of a rotation matrix
is its inverse).
Note finally that $R$ and $S$ are actually (real) {\em unitary}
matrices and hence can be viewed as {\em quantum operations} in
$2n$ dimensions i.e.~on $O(\log n)$ rebits. Here and henceforth we
will use the term ``rebit'' to refer to a qubit whose state
components (in the fixed basis being used) are restricted to be
real numbers.

The general case of $\langle Z_k\rangle$ and a general input state
$\ket{\psi_{\rm in}}$ is similar: in eq.~(\ref{Z1rebit}) we
replace $\bra{2}$ and $\ket{1}$ by $\bra{2k}$ and $\ket{2k-1}$
respectively and $S$, depending on the input state, is replaced by
the real antisymmetric matrix
\begin{equation}\label{sin}
  S[l,j] = \bra{\psi_{\rm in}} -i c_j c_l \ket{\psi_{\rm in}}.
\end{equation}
A computational basis input state $\ket{x}=\ket{x_1\cdots x_n}$
gives
\[ S(x) = \sum_{j = 1}^{2n}
(-1)^{j + x_{\lceil j /2 \rceil}}
\ket{j - (-1)^j}\bra{\rule{0mm}{3.5mm}j}
\]
which is a ``pairwise swap'' operation with a conditional
phase $(-1)^{j + x_{\lceil j/2 \rceil}}$, and then
\[
\langle Z_k(x) \rangle = \bra{2k} R S(x) R^{-1} \ket{2k-1}.
\]

The nearest neighbour restriction on matchgate actions in Theorem
\ref{one} is in fact a crucial ingredient for the classical
simulability of the circuits (as indeed n.n. and next-n.n. actions
are already universal for quantum computation \cite{jm08}). Note
that n.n. matchgate actions may be extended to arbitrary line pairs
using the qubit SWAP operation on n.n. lines. However SWAP is {\em
not} a matchgate (even though SWAP $=G(I,X)$ but det$(I)\neq$ det
$X$!) A closely related bonafide matchgate is the operation
$W=G(Z,X)$, being SWAP together with ``introduction of a minus sign
when both lines are 1'' i.e.~a fermionic SWAP operation if we view
qubit labels 0 and 1 as occupation numbers for fermionic modes. For
n.n. lines $k$ and $k+1$, the corresponding Jordan-Wigner operator
pairs are interchanged under conjugation by $W$ viz. $W^\dagger
c_{2k+1} W = c_{2k-1}$ and $W^\dagger c_{2k+2} W = c_{2k}$ and
therefore $W^\dagger Z_{k+1} W = Z_k$.

A circuit $C$ with input $\ket{x_1\cdots x_n}$ and measurement on
the $k$-th line will be called {\em equivalent} to a circuit $D$
with input $\ket{y_1 \cdots y_m}$ and measurement on the $l$-th
line, if they have the same probability distribution for their
output measurements. (Strictly speaking we should here include a
precision tolerance to allow for change of choice of finite basic
gate sets, but this technical detail may be readily accommodated
and we henceforth ignore it.)

Using the properties of $W$ above  we can standardise the form of
matchgate circuits as follows, which will be convenient for later
purposes.

\begin{lemma}\label{standard}
Let $C$ be any matchgate circuit of size $N$ and width $n\leq 2N$,
with input $\ket{x_1\cdots x_n}$ and final measurement $Z_k$ on the
$k$-th line.
Then there is an equivalent matchgate circuit $D$ of size $N+O(n^2)$
and width $n$ or $n+1$, with input $\ket{0\cdots 0}$ and final
measurement $Z_1$ on the first line.
\end{lemma}

\noindent {\bf Proof:} The description of $D$ is obtained from $C$
and its input string $x_1\cdots x_n$ as follows. Let $r$ be the
Hamming weight of the input string and let $0^k$ and $1^k$ denote
length $k$ strings of 0s and 1s, respectively. If $r$ is even we
take $n$ lines for $D$ initialised to $\ket{0^n}$. We then apply
$r/2$ $G(X,X)$ n.n.~gates to the bottom $r/2$ lines pairs to
obtain $\ket{0^{n-r}1^r}$. Next using $O(n^2)$ n.n.~$W=G(Z,X)$
gates we successively swap the $r$ lower 1s into their positions
in the input string to obtain $\pm\ket{x_1\cdots x_n}$. Next apply
the circuit $C$. Finally if $C$ has final measurement on line $k$,
apply a ladder of $k-1$ n.n.~$W$ gates to swap the $k$-th line
into the first position, and the equivalent final measurement is
now $Z_1$. This completes the description of $D$ when $r$ is even.
If $r$ is odd we take $n+1$ lines for $D$, applying $\lceil r/2
\rceil$ n.n.~$G(X,X)$ gates to get $\ket{0^{n-r}1^{r+1}}$. We then
leave line $n+1$ untouched by any further gates and proceed as
above for the first $n$ lines to obtain $D$.\hfill $\Box$

\section{Main results}\label{mainresult}

For the remainder of this paper, $MG = MG(n;N;x_1\cdots x_n;k)$
will denote a given matchgate circuit of width $n$, size $N$,
input $\ket{x_1\cdots x_n}$ and final measurement $Z_k$ on the
$k$-th line. Similarly, $QC = QC(m;M;y_1\cdots y_m)$ will denote a
quantum circuit of width $m$, size $M$,  input $\ket{y_1\cdots
y_m}$ and final measurement (without loss of generality) $Z_1$ on
the first line.

To express our main results we will need the notion of a classical
description or {\em encoding} of a quantum (matchgate or
conventional) circuit. We will assume that all our circuits are
composed of gates from fixed finite sets of one- and two-qubit
gates, whose products densely generate the class of all gates of
the kind being considered. In the proofs given in Section
\ref{proofs} below some particularly convenient sets will be
chosen but using standard arguments based on the Solovay-Kitaev
theorem \cite{nc} it may be shown that our results do not depend
on such particular choices. The action of each gate in a quantum
circuit of width $n$ and size $N$ can be specified by a triple
$(g,i,j)$ where $g$ encodes which of the finite gate types it is
and $i,j\in \{ 1,\cdots ,n\}$ specify which of the qubit lines it
acts upon. In any circuit we discard lines on which no gates act
so $N$ is at least $n/2$. The full circuit is then encoded as a
concatenation of its sequential gate applications:
\begin{equation} \label{eq:encoding}
  (g_1,i_1,j_1)(g_2,i_2,j_2)\cdots(g_N,i_N,j_N).
\end{equation}
All of these symbols may be represented as binary strings (of
constant length for the gate types $g$ and length $\log n$ for the
line numbers $i$ and $j$) giving an encoding of length $O(N\log
n)$ for any circuit of width $n$ and size $N$. This encoding may
also include specification of an input $x_1\cdots x_n$ and line
number $k$ for the final measurement, if desired. In the case of
matchgate circuits we could omit all the $j$ values since for
matchgates we always have $j_t=i_t+1$ for $t=1,\cdots ,N$.

\begin{theorem} \label{main}
  The following equivalence between matchgate circuits and general
  quantum circuits holds.
  \begin{enumerate}
    \item[(a)]
      Given any matchgate circuit $MG(n;N;x_1\cdots x_n;k)$ there
      exists an equivalent quantum circuit $QC(m;M;0\cdots0)$ with
      $m = \lceil \log n \rceil + 3$ and $M = O(N \log n)$.
      Moreover, the encoding of the circuit $QC$ can be computed
      from the encoding of the matchgate circuit $MG$ by means of a
      (classical) space\footnote{
        See the Appendix for definitions of space-bounded computation.}
      $O(\log n)$ computation.

    \item[(b)]
      Conversely, given any quantum circuit $QC(m;M;y_1\cdots y_m)$
      there exists an equivalent matchgate circuit $MG(n;N;0\cdots 0;1)$
      with $n = 2^{m+1}$ and $N = O(M 2^{2m})$.
      Moreover, the encoding of the matchgate circuit $MG$ can be
      computed from the encoding of the circuit $QC$ by means
      of a classical space $O(m)$ computation.
  \end{enumerate}
\end{theorem}

Before giving the proof in the next section, we discuss here some
consequences. First, to illustrate the content of the theorem,
consider a polynomial-sized matchgate circuit, i.e.~one with
$N=O(\poly (n))$. Then Theorem \ref{main}(a) states that it can be
simulated by a circuit $QC$ of $\poly(n)$ size and {\em
exponentially compressed width} $O(\log n)$, and the translation of
descriptions can be carried out with a comparatively modest
(log-space) computational cost. The latter remark is important in
that we would like the simulation of the matchgate circuit to be a
feature of the $QC$ circuit's computational power rather than being
subsumed in the computational power of the translation of $MG$ to
$QC$. Conversely, Theorem \ref{main}(b) with parameters $m=O(\log
n)$ and $M=O(\poly (n))$ asserts that any $\poly(n)$ sized quantum
circuit of logarithmic width $O(\log n)$ may be simulated by a
matchgate circuit of polynomial size (and hence also of polynomial
width). In this sense we have a full equivalence between the
computational power of polynomial-sized matchgate circuits and
universal logarithmic-width polynomial-sized unitary quantum
circuits (modulo classical log-space translations).

To elevate such observations to formal statements about quantum
computational complexity classes we need to introduce suitably
defined families of computational tasks (parameterised by
increasing input size $n$) and associated computational complexity
classes.
When characterising computational complexity in terms of the circuit
model, we need to consider families of circuits subject to a suitable
restriction for how these circuits are generated as a function of the
input and its size $n$.
In contrast, the Turing machine (TM) model requires no such auxiliary
condition, as a single TM can deal with inputs of all lengths.
For example, for usual polynomial-time quantum computation we
conventionally use families of circuits whose descriptions are
generated from the input string by a polynomial-time classical
computation.
But the complexity classes that we are aiming to characterise
(viz. poly-sized matchgate computations and a notion of log-space
quantum computation) are themselves no stronger than classical
polynomial-time computation, so we need to adopt a more strict
condition.
A natural choice is classical log-space generation, commonly used in
classical complexity theory for classes that are contained within
classical polynomial-time computation.

Let $\Sigma^*$ denote the set of all finite length bit strings.
A {\em promise problem} is a pair $A = (A_\yes, A_\no)$ of subsets
$A_\yes,A_\no\subseteq\Sigma^{\ast}$ with
$A_\yes \cap A_\no = \varnothing$.
Intuitively speaking, a promise problem expresses a computational
decision problem, where the strings in the set $A_\yes$ are those
input strings for which the correct answer is ``yes'' (or the binary
value~1) and the strings in $A_\no$ are those for which the correct
answer is ``no'' (or the binary value~0).
Input strings contained in neither of the sets $A_\yes$ and $A_\no$
are ``don't care'' inputs for which no answer is required.
Promise problems for which $A_\yes\cup A_\no = \Sigma^{\ast}$,
i.e.~that disallow ``don't care'' inputs, are called
{\em languages} (and are generally represented simply by the set
$A_\yes$).

With this terminology in mind, consider the following definitions.

\begin{definition}\label{firstdef}
Log-space generated families of matchgate circuits and the complexity
classes $\class{BMG}$, $\class{PMG}$, $\class{BQL}_\class{U}$ and
$\class{PQL}_\class{U}$ are defined as follows.
\begin{enumerate}
\item[(a)]
  A {\em log-space generated} family of matchgate circuits is a
  classical log-space computable function $g$ on $\Sigma^*$ such that,
  for each $w \in \Sigma^*$, the string $g(w)$ is an encoding of a
  matchgate circuit $MG(n; N; x_1\cdots x_n;k)$.

\item[(b)]
  A promise problem $A = (A_\yes,A_\no)$ is computed with bounded
  error by a log-space generated family $g$ of matchgate circuits if and
  only if, for all strings $w \in A_\yes$ the matchgate circuit
  encoded by $g(w)$ outputs 1 with probability at least $2/3$, and for
  all strings $w \in A_\no$ the matchgate circuit encoded by $g(w)$
  outputs 1 with probability at most $1/3$.

\item[(c)]
  A promise problem $A = (A_\yes,A_\no)$ is computed with unbounded
  error by a log-space generated family $g$ of matchgate circuits if and
  only if, for all strings $w \in A_\yes$ the matchgate circuit
  encoded by $g(w)$ outputs 1 with probability strictly greater than
  $1/2$, and for all strings $w \in A_\no$ the matchgate circuit
  encoded by $g(w)$ outputs 1 with probability at most $1/2$.

\item[(d)]
  The class $\class{BMG}$ consists of all promise problems
  computable with bounded error by a log-space generated family $g$ of
  matchgate circuits, and the class $\class{PMG}$ consists of all
  promise problems computable with unbounded error log-space generated
  family $g$ of matchgate circuits.

\item[(e)]
  The class $\class{BQL}_\class{U}$ consists of all promise problems
  computable with bounded error by a log-space unitary quantum
  computation, and the class $\class{PQL}_\class{U}$ consists of all
  promise problems computable with unbounded error by a log-space
  unitary quantum computation.
  (These notions are described in the Appendix.)
\end{enumerate}
\end{definition}

\noindent It was proved in refs.~\cite{wat1,wat2} that, when
quantum operations are restricted to those expressible by matrices
of algebraic numbers, it holds that $\class{PQL}_\class{U} =
\class{PL}$, where $\class{PL}$ is the class of promise problems
computed with unbounded error by classical probabilistic Turing
machines operating in logarithmic space.

In terms of the above defined notions, Theorem \ref{main} has the
following immediate corollaries.

\begin{corollary}\label{cor1}
  Let $\cp = \{(p_0^w,p_1^w): w\in\Sigma^*\}$ be a family of
  probability distributions on Boolean values, parameterised by binary
  strings $w$.
  Then $\cp$ arises as the output distribution for a log-space
  generated family of matchgate circuits if and only if $\cp$ arises
  as the output distribution for a unitary log-space quantum
  computation.
\end{corollary}

\noindent {\bf Proof.} The two directions of implication follow
easily from Theorem \ref{main}, noting that then both directions
of translation are computable in classical log-space, and using
the interpretation (mentioned in the Appendix, taking $f(n)=\log
n$ there) of log-space quantum computation in terms of log-space
generated families of log-width quantum circuits. \hfill\qed

\begin{corollary}\label{cor2}
  It holds that
  \begin{enumerate}
  \item[(a)]
    $\class{BMG} = \class{BQL}_\class{U}$, and
  \item[(b)]
    $\class{PMG} = \class{PQL}_\class{U}$.
  \end{enumerate}
\end{corollary}

We may alternatively use the notion of {\em completeness} of a
language or promise problem for a complexity class to express the
equivalence of the computational power of polynomial-sized matchgate
circuits and log-space quantum computation.
A promise problem $A = (A_\yes,A_\no)$ is {\em complete}
for a complexity class $\cc$ if $A \in \cc$ and any promise problem
$B = (B_\yes,B_\no)$ in $\cc$ can be {\em reduced} to $A$.
There are many differing notions for the reduction of one problem to
another, but we will consider {\em log-space reductions}.
The choice to consider log-space reductions is analogous to the
log-space generation of circuit families, and provides a
non-trivial notion of completeness for the classes under discussion.

\begin{definition}\label{defred}
  A function $g:\Sigma^* \rightarrow \Sigma^*$ is a
  {\em log-space reduction} from a promise problem $A=(A_\yes,A_\no)$
  to another promise problem $B = (B_\yes,B_\no)$ if $g$ is a
  (classical) log-space computable function and it holds that
  $g(w) \in B_\yes$ for all $w\in A_\yes$, and
  $g(w) \in B_\no$ for all $w\in A_\no$.
\end{definition}

\noindent
Thus, a promise problem $B$ is {\em complete} for
$\class{BQL}_\class{U}$ if $B\in\class{BQL}_\class{U}$ and, for every
promise problem $A\in\class{BQL}_\class{U}$, there exists a log-space
reduction $g$ from $A$ to $B$.
The definition for $\class{PQL}_\class{U}$ is similar.

Now consider the following promise problems that formalize the
computational tasks of simulating matchgate circuits of the sort that
we consider.
(These problems are stated in a standard way that makes clear which
strings belong to the yes- and no-sets.)

\begin{trivlist}
\item
  \textsc{The bounded-error matchgate circuit problem} ($\text{BMGC}$
  for short)
\item
  \begin{tabular*}{\textwidth}{@{}l@{\extracolsep{\fill}}p{5.25in}@{}}
    {\it Input:} & An encoding of $MG(n;N;x_1\cdots x_n;k)$, a nearest
    neighbour quantum matchgate circuit, its input string, and its output
    qubit.\\[2mm]
    {\it Yes:} &
    The output of $MG$ is 1 with probability $p \geq \frac{2}{3}$.
    \\[2mm]
    {\it No:} &
    The output of $MG$ is 1 with probability $p \leq \frac{1}{3}$.
  \end{tabular*}
\end{trivlist}

\begin{trivlist}
\item
  \textsc{The unbounded-error matchgate circuit problem} ($\text{MGC}$
  for short)
\item
  \begin{tabular*}{\textwidth}{@{}l@{\extracolsep{\fill}}p{5.25in}@{}}
    {\it Input:} & An encoding of $MG(n;N;x_1\cdots x_n;k)$, a nearest
    neighbour quantum matchgate circuit, its input string, and its output
    qubit.\\[2mm]
    {\it Yes:} &
    The output of $MG$ is 1 with probability $p > \frac{1}{2}$.
    \\[2mm]
    {\it No:} &
    The output of $MG$ is 1 with probability $p \leq \frac{1}{2}$.
  \end{tabular*}
\end{trivlist}

\noindent The equivalence between matchgate circuits and
space-bounded quantum computations established by
Theorem~\ref{main} gives the following corollary.

\begin{corollary}\label{cor3}
  With respect to classical log-space reductions, we have
  \begin{enumerate}
  \item[(a)]
    $\textup{BMGC}$ is $\class{BQL}_\class{U}$-complete, and
  \item[(b)]
    $\textup{MGC}$ is $\class{PQL}_\class{U}$-complete.
  \end{enumerate}
\end{corollary}

Thus intuitively, any instance of any problem in
$\class{BQL}_\class{U}$ can be reduced to approximating the output of
a matchgate circuit having bounded error, and conversely the latter
can be computed in $\class{BQL}_\class{U}$.
This corollary is formally similar to the result of \cite{AG}
characterising the computational power of stabiliser circuits, where
it is shown that an analogously defined language for stabiliser
circuits (rather than our matchgate circuits here) is complete for
the complexity class $\oplus$L (again relative to log-space
reductions).

\section{Proof of Theorem \ref{main}}\label{proofs}

\subsection{Proof of Theorem \ref{main}(a)}

We will first describe a procedure for converting $MG(n;N;x_1\cdots
x_n;k)$ into a circuit $QC$ of the claimed size and width, and then
discuss the implementation of this procedure in space $O(\log n)$.
We will consider first the case where the given matchgate circuit
takes the form $MG(n;N;0\cdots 0;1)$, and then apply Lemma
\ref{standard} to handle the general case. Without loss of
generality, it will be assumed that $n$ is a power of~2, and
hereafter we will let $m = \log(n) + 3$.

For clarity of exposition (and without loss of generality) we will
assume that all our matchgates are drawn from a particular finite
set chosen as follows. Recall from eq (\ref{mgrot}) that
matchgates correspond surjectively with $SO(4,\RR )$ rotations.
Using the standard Euler decomposition \cite{euler} any such
4-dimensional rotation can be expressed as a product of six
rotations each of which acts nontrivially in one of the six
2-dimensional co-ordinate subspaces of $\RR^4$. Accordingly we
choose six basic matchgates that correspond to six such rotations,
through an angle that is an irrational multiple of $\pi$. For
example we may simply take the rotations to be
\begin{equation}\label{arctan} \frac{1}{5} \left( \begin{array}{cc} 4 & -3
\\ 3 & 4
\end{array} \right) \end{equation}
with rotation angle $\arctan 3/4$. Then any matchgate (to suitable
precision) may be represented as a circuit of these six basic
matchgates. For convenience (cf Lemma \ref{standard}) we can also
include $G(X,X)$ and $G(Z,X)$ in our basic set.

Recall (cf eq.~(\ref{Z1rebit})) that the output probabilities of the
matchgate circuit satisfy
\[
p_0 - p_1 = \langle Z_1\rangle =\bra{2} RSR^{-1} \ket{1} = \bra{1}
S^{-1}RSR^{-1}\ket{1}.
\]
Here $R = R_N \cdots R_1$ is the $2n$-dimensional rotation
corresponding to the product of the individual two-level rotations
$R_t$ of the matchgates $(g_t,i_t,i_t+1)$ in the circuit $MG$, and
$S$ is the $2n$-dimensional matrix of eq.~(\ref{ess}). The circuit
$QC$ will be constructed to implement the computation illustrated
in Figure 1  for the unitary $U=S^{-1}RSR^{-1}$.

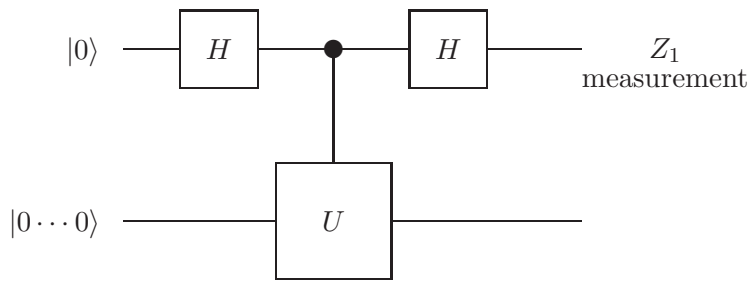
\begin{figure}[!ht]
\begin{center}
  \setlength{\unitlength}{1in}
  \begin{picture}(6,1.5)(-0.75,0.45)
    \put(0.6,0.65){$\ket{0\cdots 0}$}   \put(0.9,1.55){$\ket{0}$}
    \put(3.6,1.4){\shortstack{$Z_1$\\ measurement}}
    \put(1.2,0.7){\line(1,0){0.8}}
    \put(1.2,1.6){\line(1,0){0.3}}
    \put(2.6,0.7){\line(1,0){1.0}}
    \put(1.9,1.6){\line(1,0){0.8}}
    \put(3.1,1.6){\line(1,0){0.5}}
    \put(2.3,1.0){\line(0,1){0.6}}
    \put(2.0,0.4){\framebox(0.6,0.6){$U$}}
    \put(1.5,1.4){\framebox(0.4,0.4){$H$}}
    \put(2.7,1.4){\framebox(0.4,0.4){$H$}}
    \put(2.3,1.6){\circle*{0.1}}
  \end{picture}
\end{center}
\label{figure:1}
\caption{
  The process performed in the translation of $MG$ to $QC$ involves
  $\log(2n)+1$ qubit lines all initialised to $\ket{0}$.
  The middle gate is a controlled-$U$ operation $\Lambda U$ with the top
  line as control line.
  $H$ denotes the Hadamard gate.
  The final $Z_1$ measurement on the first line then has the unique
  binary probability distribution whose mean is
  $\langle Z_1\rangle = \op{Re} \bra{0\cdots 0}U\ket{0\cdots 0}$.
  (Similar processes have been used many times in the quantum computing
  literature.)
}
\end{figure}

The real unitary operation $U=S^{-1}RSR^{-1}$ acts in $2n$
dimensions, and for the computation described in Figure~1 we need
to provide a description of the controlled operation $\Lambda U$ as
a circuit of two-qubit gates on $m$ rebit lines. Each basic
rotation $R_t$ corresponding to a matchgate $(g_t,i_t,i_t+1)$ acts
nontrivially in the span of two basis states selected from
$\ket{2i_t-1},\ket{2i_t}\ket{2i_t+1}$ and $\ket{2i_t+2}$, and it
acts as the identity in the $(2n-2)$-dimensional complementary
subspace.
Our strategy now is the following: we begin by expressing each of
these basic rotations as a constant sized circuit of
multiply-controlled single qubit rotations. Then we will use the
fact \cite{ninepaper} that any $r$-fold controlled single qubit
gate can be implemented using $O(r)$ one- and two-qubit
operations, to build up our final circuit of elementary gates, of
the claimed size.

Unfortunately, if we represent the dimensions $\ket{i}$ as rebit
lines via the usual binary representation of the labels $i=1,
\ldots ,2n$, then $R_t$ will {\em not} be a two-qubit gate, and
indeed it will generally act on all $\log (2n)$ rebit lines. To
circumvent this problem we instead use a Gray code to associate
rebit lines to the $2n$ dimensions. A Gray code \cite{graycode}
for the numbers $1, \ldots ,2n$ is a sequence of distinct binary
strings, each of length $\log(2n)$, such that the strings
corresponding to any consecutive numbers $i$ and $i+1$ have
Hamming distance 1. Such codes exist and the Gray code strings can
be easily computed from the standard binary representation via a
simple sequence of single bit additions \cite{graycode}.

Now, as noted above, each basic matchgate rotation $R_t$ acts within
two of four consecutive dimensions and hence within the space of
two Gray code strings of Hamming distance at most 3.
If we conjugate $R_t$ by suitable $X$ and $\Lambda X$ gates that act on
the rebit lines where the Gray code strings differ, and that are
controlled by the Gray code line values where they agree, then the
Hamming distance may always be reduced to 1, i.e.~each $R_t$ (with
the conjugations in place) can be made to act within the
2-dimensional subspace of a single rebit line conditioned on the
bit values of all other lines that are specified by the (common)
Gray code values on those lines.

As an illustrative example, suppose $R_t$ acts within the
2-dimensional span of the computational basis states
$\ket{0 1 z_3 1 z_5 \cdots z_{m-2}}$
and $\ket{1 0 z_3 0 z_5\cdots z_{m-2}}$ having Hamming distance~3,
where $z_3 z_5 \cdots z_{m-2}$ is the fixed substring common to
the two Gray code strings.
If we apply the gate $X_2$, then $X_4$, then $(\Lambda X)_{12}$, and
finally $(\Lambda X)_{14}$, each conditioned on the values
$z_3 z_5 \cdots z_{m-2}$ of lines 3 and 5 to $m-2$,
then we obtain the basis states $\ket{ 0 0 z_3 0 z_5 \cdots z_{m-2}}$
and $\ket{1 0 z_3 0 z_5 \cdots z_{m-2}}$, respectively, i.e.~the
rotation action is mapped into the 2-dimensional subspace of the
first rebit controlled by the values $0 z_3 0 z_5\cdots z_{m-2}$
on the remaining lines.

Next we use the algorithm from Section~7.5 of \cite{ninepaper} for
representing an $r$-fold controlled operation $\Lambda^r T$ (for a
single qubit gate $T$) in terms of a circuit of $O(r)$ two-qubit
gates, which requires the addition of one additional ancillary
qubit. Applying this for $r\leq m-2$, and where $T$ is our
2-dimensional rotation or one of the extra $X$ operations in the
conjugations above, we obtain the description of the controlled-$R$
operation (and analogously also controlled-$R^{-1}$) needed for the
computation in Figure~\ref{figure:1} as a circuit of two-qubit
gates.

Similarly, for the operation $S$, we see from its block diagonal
form eq.~(\ref{ess}) that $S$ can be represented as a product of
operations each of which acts non-trivially only in 2 consecutive
dimensions. Hence using the same techniques as above we can
express $S$ as a circuit of 2-local gates on rebit lines
corresponding to the Gray code. Alternatively we can note that
eqs. (\ref{ess},\ref{eqy}) show that in the usual binary
representation of the basis state labels, $S$ has the local form
$S=\tilde{Y}_1$.
Then using a simple $O(\log n)$ sized circuit
\cite{graycode} that translates between the Gray code and usual
binary representation, we can get a circuit of size $O(\log n)$
for the required controlled-$S$ operations.

Assembling these ingredients, we obtain a circuit on $m=\log(2n)+2$
rebit lines described in Figure~\ref{figure:1} for the operation
$U=S^{-1}RSR^{-1}$.
If the original matchgate circuit had size $N$ and width $n$ then the
final circuit has size $O(N \log n)$ and width $m$.
The $O(\log n)$-factor increase in size arises from the algorithm of
\cite{ninepaper} to decompose an $O(\log n)$-controlled operation into
elementary operations.
Thus we have achieved the transformation
\[
MG(n;N;0\cdots 0;1) \rightarrow QC(m; O(N \log n); 0\cdots 0).
\]

For the general case where $MG$ takes the form $MG(n;N;x_1\cdots
x_n;k)$, one may follow the same procedure above after first
applying the translation process of Lemma \ref{standard}. This
requires the concatenation of sequences of gates to the beginning
and end of the encoding of the circuit $MG$, but otherwise does
not involve any computation on the gates of $MG$ itself. This
results in a matchgate circuit of the standard form $MG(n;N +
O(n^2);0\cdots 0;1)$.

Finally we point out that the succession of translations described
above can be achieved by a classical computation bounded to take place
in $O(\log n)$ space by noting the following facts:
\begin{enumerate}
\item[(i)]
The gates of $MG$ are processed one at a time to obtain the gates of
$QC$.
The computation requires one pass through the encoding of $MG$
to construct the gates of $QC$ corresponding to the operator $R$, and
a second pass in the opposite direction to obtain the gates of $QC$
corresponding to $R^{-1}$.

\item[(ii)]
Each gate of $MG$ is encoded as a string of length $O(\log n)$, and
it is clear that the computation of the gates of $QC$ corresponding to
each such gate can be computed in space $O(\log n)$.
In particular, space $O(\log n)$ is sufficient to translate any number
$i\in\{1,\ldots,2n\}$ expressed in binary into its Gray code
representation, to compute the required conjugations by $X$ and
$\Lambda X$ gates, and to implement the procedure of \cite{ninepaper}
on the required gates
(keeping in mind that the individual two-qubit gates
resulting from this procedure are output sequentially and need not all
be stored simultaneously).

\item[(iii)] The additional gates that are added by the
construction of Lemma~1 are easily generated one-by-one in space
$O(\log n)$, and can be incorporated into the procedure described
above by the addition of $O(\log n)$ space.
\end{enumerate}

\noindent This completes the proof Theorem \ref{main}(a).

\subsection{Proof of Theorem \ref{main}(b)}

Given a description of a quantum circuit $QC(m;M;y_1\cdots y_m)$
we will generate an equivalent matchgate circuit
$MG(n=2^{m+1};N=O(2^{2m}M);x_1\cdots x_n=0\ldots 0;k=1)$ and the
translation of description will be computable in space $O(m)$.
Without loss of generality we may take $y_1\cdots y_m=0\cdots 0$
(as the description of the $QC$ circuit may be easily prefixed by
$X$ gates corresponding to the given $y_1\cdots y_m$ values.)

Let $V$ denote the total unitary operation of the $QC$ circuit.
Then its output distribution is the unique binary distribution
with mean
\begin{equation}\label{z1qc}
  \langle Z_1\rangle_{QC} =
  \bra{0\cdots 0}V^\dagger Z_1 V \ket{0\cdots 0}.
\end{equation}
On the other hand, for a hypothetical matchgate circuit $MG$ of width
$n$ we have
\[
\langle Z_1 \rangle_{MG} = \bra{2} RSR^{-1} \ket{1}
\]
where $R,S\in SO(2n,\RR)$ arise from the circuit as described in
Section \ref{matchsec}. If we represent the $2n$ dimensions in
binary as rebit lines then eq.~(\ref{eqy}) shows that
$S=\tilde{Y}_1$ and
\begin{equation}\label{z1mg1}
  \langle Z_1 \rangle_{MG} = \bra{2} R\tilde{Y}_1R^{-1} \ket{1}.
\end{equation}
We will define $MG$ by exploiting the structural similarity
between eqs.~(\ref{z1qc}) and (\ref{z1mg1}). Roughly speaking we
will associate matchgates to the successive gates of $V$, choosing
them so that their adjoint actions as per eq.~(\ref{mgrot}) will
give an overall rotation that reproduces $V^\dagger$. In this
metamorphosis of eq.~(\ref{z1qc}) into eq.~(\ref{z1mg1}) we will
need to address three discrepancies between these formulae: (i)
$V$ is generally complex whereas $R$ is real; (ii) $\langle
Z_1\rangle_{MG}$ is expressed as an off-diagonal matrix element
whereas $\langle Z_1\rangle_{QC}$ is a diagonal element; and (iii)
$\langle Z_1\rangle_{QC}$ has a central $Z_1$ term whereas
$\langle Z_1\rangle_{MG}$ has $\tilde{Y}_1$ in this position.

We begin by treating (ii) and (iii) together.
Introduce the two-qubit controlled operation
\[
v=\proj{+} \otimes I+\proj{-}\otimes \tilde{Y}
\]
where $\ket{\pm}= \frac{1}{\sqrt{2}}(\ket{0}\pm \ket{1})$ are
$X$--basis states. Then
\begin{equation}\label{veq}
  v^\dagger (\tilde{Y}_1\otimes I_2)v
  = Z_1\otimes (-\tilde{Y}_2).
\end{equation}
Next introduce the qubit swap operation $SWAP_{12}$ and
$w_{12}=SWAP_{12}\, v$.
Then eq.~(\ref{veq}) gives
\begin{equation}\label{weq}
  w^\dagger (I_1 \otimes \tilde{Y}_2) w
  = Z_1\otimes (-\tilde{Y}_2).
\end{equation}
Now we extend the $QC$ circuit by introducing an ancillary qubit $A$
(hereafter located in the rightmost position in the binary description
of the computational basis) and a post-processing by $w_{1A}$ (where 1
denotes the first line of the given $QC$ circuit) to obtain
$\tilde{V}=w_{1A}V$.
Then eq.~(\ref{weq}) (with label 2 in that equation replaced by label
A) gives
\begin{equation}\label{vave}
  \begin{array}{rcl}
    \bra{0\cdots 0 1_A}\tilde{V}^\dagger \tilde{Y}_A \tilde{V}
    \ket{0\cdots 00_A} &  =  & \bra{0\cdots 0}V^\dagger Z_1 V
    \ket{0\cdots 0} \bra{1_A}-\tilde{Y}_A\ket{0_A}\\ & = &
    \bra{0\cdots 0}V^\dagger Z_1 V \ket{0\cdots 0}.
  \end{array}
\end{equation}
Note that if we label our Hilbert space dimensions by natural
numbers counting from 1 to $2^{m+1}$ then $\ket{0\cdots 00_A}$ and
$\ket{0\cdots 01_A}$ correspond to $\ket{1}$ and $\ket{2}$
respectively and eq.~(\ref{vave}) gives
\[
\langle Z_1\rangle_{QC}
= \bra{2} \tilde{V}^\dagger \tilde{Y}_A \tilde{V} \ket{1}.
\]
Comparing this expression with $\langle Z_1\rangle_{MG}$ in
eq.~(\ref{z1mg1}) we see that the off-diagonal matrix elements as well
as the central $\tilde{Y}$-terms (acting on the first line in both
cases) are now in correspondence thus addressing issues (ii) and (iii)
above.
Furthermore the translation of the $QC$ circuit description for $V$
into that for $\tilde{V}$ (i.e.~adjoining a description of $w_{1A}$)
is a simple process that can be carried out even in constant space.

To deal with issue (i) recall that real gates suffice for
universal quantum computation \cite{bv} and given any quantum
circuit there is an equivalent circuit comprising real gates from
the special orthogonal group that has one extra ancillary rebit line
$B$. This construction is a direct consequence of the algebra
isomorphism
\[ a+ib \mapsto aI-b\tilde{Y} = \left( \begin{array}{cc} a & -b \\
b & a \end{array} \right) \] between complex numbers and a class
of real $2 \times 2$ matrices. Correspondingly to any $K$-qubit
unitary gate $U\in U(2^K)$ we associate the real $SO(2^{K+1},\RR)$
gate
\[
\hat{U} = \op{Re}(U) \otimes I_B - \op{Im}(U) \otimes \tilde{Y}_B
\]
and to any $K$-qubit state $\ket{\psi}$ we associate the
$(K+1)$-rebit state $|\tilde{\psi}\rangle = \op{Re}(\ket{\psi})
\otimes \ket{0}_B + \op{Im}(\ket{\psi}) \otimes \ket{1}_B$ to map
any quantum computation into an equivalent real one.

If the gates $U$ of the $QC$ circuit are from a universal set of
one- and two-qubit gates then the corresponding $\hat{U}$'s will
be one-, two-, and three-rebit gates and we may assume that they
are given already decomposed in terms of a finite real universal
set of two-rebit gates. For this set we choose the six rotations
through angle $\arctan 3/4$ (cf eq.~(\ref{arctan})) that act
respectively in the six 2-dimensional co-ordinate subspaces of
$\RR^4$. In this way we translate the original $QC$ circuit into
an equivalent circuit of width $m+2$ comprising these basic real
gates and the translation can be achieved in $O(m)$ space, for
example by sequentially translating each successive gate of
$\tilde{V}$ into its real version.

Each of the resulting basic gates is a rotation $T_{k,k'}$ acting
on some (not necessarily nearest neighbour) pair $(k,k')$ of rebit
lines. Let $\tau$ be the associated $SO(4,\RR)$ matrix, which by
our choice of basic gates, acts non-trivially only in a
2-dimensional subspace of $\RR^4$. To associate a circuit of
matchgates on $n=2^{m+1}$ qubit lines we need to view the global
gate $T= T_{k,k'}\otimes_{j\neq k,k'} I_j$ as an operator on the
full $2n=2^{m+2}$-dimensional space with basis labelled {\em
sequentially} as $\ket{1},\ldots ,\ket{2n}$. By virtue of its
tensor product structure the $2n\times 2n$ matrix of $T$ amounts
to the application of $n/2$ parallel instances of $\tau$, acting
in a set of $n/2$ disjoint 4-dimensional co-ordinate subspaces.
Furthermore $\tau$ acts non-trivially in only 2 of its 4
dimensions so $T=T_1T_2\cdots T_{n/2}$ is then given as a product
of $n/2$ rotations $T_i$, in $2n$ dimensions, each of which acts
non-trivially only in the span of two basis states $\ket{k_i}$ and
$\ket{l_i}$ (and acts as the identity in the $2n-2$ dimensional
complement). Now if $|k_i-l_i|\leq 3$ then the rotation $T_i$ is
induced directly by a single matchgate $G_i$ via
eq.~(\ref{mgrot}). Here $G_i$ is one of six basic matchgates that
correspond to our choice of the six basic rotations. If
$|k_i-l_i|> 3$ recall that the matchgate $G(Z,X)$ on qubit lines
$k,k+1$ induces an associated rotation that swaps the pairs
$(2k-1,2k)$ and $(2k+1,2k+2)$ with each other. Then the rotation
$T_i$ is induced by a basic matchgate $G_i$ conjugated by a ladder
of $O(n)$ $G(Z,X)$ matchgates (to bring $(k_i,l_i)$ into a
position with $|k_i-l_i|\leq 3$). Thus each $T_i$ corresponds to a
circuit of $O(n)$ matchgates so $T$ (comprising $O(n)$ $T_i$'s)
corresponds to a matchgate circuit of size $O(n^2)$. Then finally,
our original $QC$ circuit of width $m$ and size $M$ is translated
into an equivalent matchgate circuit of width $2^{m+1}$ and size
$O(Mn^2) = O(M2^{2m})$ as claimed. Note that in the final
translation, the renaming of the six basic rotations by their
corresponding matchgates is a simple operation but the
corresponding line labels for the matchgate actions range from 1
to $n=2^{m+1}$ needing $O(m)$-bit specifications, which can be
calculated in $O(m)$ space using standard arithmetic operations.
This completes the proof of Theorem \ref{main}(b).

\section{Concluding remarks}
\label{ending}

We have demonstrated an equivalence (in the precise sense given in
Section \ref{mainresult}) between the computational power of
matchgate circuits of width $n$ and universal unitary quantum
computation within an exponentially compressed space bound of
$O(\log n)$ qubits. This equivalence is particularly interesting
in the case of classically efficiently simulatable quantum
computations (cf Theorem \ref{one}) where the matchgate circuits
(and then also the corresponding log-space bounded circuits) have
$\poly (n)$ size. In the literature there are two seemingly
different techniques for constructing classically simulatable
quantum computations: (a) we have the normaliser formalism that
underlies the simulability of stabiliser circuits, as expressed in
the Gottesman-Knill theorem \cite{cliffgp1}. Generalisations of
this technique have been considered in \cite{cjl} and it has been
argued in \cite{j2} that the classical simulability of matchgate
circuits in Theorem \ref{one} can be understood in terms of a
suitably generalised normaliser construction too.
Characteristically this technique leads to circuits of gates that
are subject to algebraic constraints (normaliser conditions) and
despite their classical simulability, the associated computations
can generate complex entangled states. (b) A second method for
imposing classical simulability is to restrict the amount of
entanglement in the states that may be generated in the course of
the computation. Such states have correspondingly small classical
descriptions which can then be exploited for classical simulation.
Examples of this method include the near-separable computations of
\cite{jlin,vid} and use of the matrix product formalism (cf
\cite{vcm} for a comprehensive recent review) to impose bounds on
the Schmidt rank across any partition of a  multi-qubit state.

With the above in mind, our main results can be seen as providing
a link between (a) and (b) showing that, at least for the case of
matchgate computations, the associated normaliser formalism may be
viewed as an example of a limitation on the amount of generic
entanglement and {\em vice versa}.

Our results (especially Corollary \ref{cor2}) provide alternative
characterisations of some classically simulatable quantum
complexity classes. Such classes, {\em ipso facto} contained
within \class{P} (classical polynomial time computation), have the
following interesting feature: suppose we have any class \class{C}
of classically simulatable quantum computations (that is not just
classical computation itself i.e. we have a nontrivial quantum
gate involved). Then we can argue that if the computational power
of \class{C} is {\em full} classical polynomial time (in a
suitably strong sense, as below), it would follow that \class{BPP}
equals \class{BQP} i.e. that the powers of polynomial time
classical and quantum (bounded error) computing coincide. Thus it
is significant that our poly-sized circuits of matchgates should
have a computational power that is strictly weaker than \class{P}
(and similarly for stabiliser circuits). The intuitive argument is
the following: we interpret the condition that \class{C} has full
classical poly-time power in a strong sense, as meaning that any
classical gate can be simulated (in a quantum-coherent fashion) by
a circuit from \class{C}. Thus we can simulate a Toffoli gate. But
it is known \cite{shi} that the Toffoli gate plus any nontrivial
(basis-changing) quantum gate is efficiently universal for quantum
computation. Hence if \class{C} also has any such nontrivial
quantum gate, then in addition to being classically simulatable,
\class{C} must also be quantum universal. Note that the above
argument does not apply to some possible classes of classically
simulatable computations such as those in \cite{vdn}, that can
involve further global restrictions on the structure of allowed
circuits e.g. although it may be possible to simulate a Toffoli
gate, the definition of the class may yet exclude its appearance
in a general position within a circuit.

Finally we mention some possible avenues of significance of our
results for further physical and implementational considerations.

Matchgate circuits are known to include the real-time dynamics of
one-dimensional XY spin chain Hamiltonians with nearest neighbor
interactions (cf. e.g., \cite{terdiv,jm08}). Thus Theorem
\ref{main}(a) implies that a digital {\em quantum} simulation of the latter
can be carried out employing a quantum computer which is exponentially
smaller in its use of qubits than the original system. This may
for example, allow for an experimental observation of a quantum
phase transition induced by varying the parameters in the
Hamiltonian (cf. \cite{vcl08}).

If there exist experimental situations in which matchgates are
especially easy to realise, then the simulation of universal
quantum computation using matchgates, that is implied by Theorem
\ref{main}(b), may be of experimental interest to demonstrate
quantum computational effects, albeit with an exponential overhead
in circuit width. Recalling that matchgates correspond to unitary
evolutions generated by quadratic Hamiltonians of fermionic
creation and annihilation operators \cite{terdiv,jm08}, a
prospective setting here might be a fermionic counterpart
\cite{bdek04} of the Knill-Laflamme-Milburn scheme in linear
optics, where the gates are supposed to be implemented by
beam splitters using the point contacts of a two-dimensional
electron gas.

Lastly we recall that the symmetry of the special orthogonal group
appeared abstractly in our developments, via the Clifford algebra
formalism (cf eq. \ref{mgrot}). However it can also be
emergent {\em  physically} as exotic
quantum statistics. For example, in the fractional  quantum Hall effect
with the filling fraction $\nu = 5/2$ or in recent
topological insulator-superconductor structures, the
quasi-particle  excitations in the celebrated Moore-Read state
behave as non-abelian (Ising) anyons \cite{mr91}, that can
realise non-abelian quantum  statistics of $SO(2n, \RR)$ for $2n$
excitations \cite{nw96}, analogous to the transformations available in the
matchgate formalism. The experimental observation of such anyonic
behaviour is of current interest for condensed-matter physics, as well as
for a potential implementation of topological quantum
computation, where the anyon braiding operations would need to be
supplemented by further non-topological operations to achieve
universal quantum computation in the full Hilbert space
\cite{tqc}. However Theorem~\ref{main}(b) implies that the
$SO(2n, \RR)$ action has an inherent universal
computational capability, albeit within an exponentially
compressed quantum space bound, that can then also be viewed in
terms of matchgate computations.

\appendix
\section{Appendix: Definitions of space-bounded computation}
\label{appendix}

The notion of classical space-bounded computation that we use is
the standard textbook one (cf \cite{papadim,AB}), and for
completeness we include a summary description here. The classical
Turing machines (TMs) that we consider have three tapes called the
input, work, and output tapes, respectively. Each tape is a
sequence of cells which may contain one of three symbols: the
blank symbol B, or one of the binary values 0 or 1. (Often one
allows more symbols, but these ones are sufficient for the present
discussion.) Conventionally we take the tapes to be infinite to
the right.

At any stage the TM has an {\em internal state} $q$, chosen from a
fixed finite set, and each tape has a {\em tape head} that scans
one of its cells at each moment. The computational process of the
TM is defined by a finite table of transition rules of the form
\begin{equation}\label{clinstr}
  (q,s_\text{input},s_\text{work})
  \rightarrow
  (r, t_\text{work},t_\text{output},d_\text{input},d_\text{work})
\end{equation}
where $q$ is an internal state and $s_\text{input},s_\text{work}$
are symbols on the input and work tape, respectively. Each $d$ is
either $L$ (``one step left''), $R$ (``one step right'') or $S$
(``stay in the same place''). The meaning of this instruction is
the following: if the TM has the internal state~$q$, and is
scanning the symbols $s_\text{input}$ and $s_\text{work}$ on its
input and work tapes, then
\begin{enumerate}
\item[(i)] the internal state changes to $r$, \item[(ii)] the
input tape head moves in direction $d_\text{input}$, \item[(iii)]
the symbol being scanned on the work tape is replaced by
  $t_\text{work}$ and the work tape head moves in direction
  $d_\text{work}$, and
\item[(iv)] if $t_\text{output}$ is a non-blank symbol, then this
  symbol is written on the output tape and the output tape head moves
  right (while nothing happens to the output tape or tape head if
  $t_\text{output}$ is a blank symbol).
\end{enumerate}
These rules imply that the input tape is ``read only'' and the
output tape is ``write only''. This is done so that the work tape
effectively represents the memory of the Turing machine, upon
which various limitations can be defined. It is also common to
impose the restriction that if $s_\text{input}$ is the blank
symbol then $d_\text{input}$ cannot be $R$, which forces the input
tape head to remain at most one square away from the actual input
string on the input tape. (If this condition is not imposed, the
position of the input tape head could be exploited to effectively
store a large integer value that is not counted as part of the
Turing machine's memory usage.) We assume the machine to be {\em
deterministic}, which means that there can be only one transition
rule whose left-hand-side is given by any triple
$(q,s_\text{input},s_\text{work})$.

To say that a TM is run on a particular binary input string $w$
means that the TM is started in a distinguished starting state
$q_{\text{start}}$, has $w$ written left-justified on the input
tape, one symbol per cell and with all other tape cells containing
the blank symbol, and with all tape heads scanning the leftmost
cells of their respective tapes. The transition rules then
determine the steps of the computation. One of the TM's internal
states $q_{\text{halt}}$ is designated as the {\em halting state},
and the computation is assumed to stop in the event that this
state is reached. If this happens, the binary string written on
the output tape is taken as the TM's output. A function
$g:\Sigma^* \rightarrow \Sigma^*$ is computed by the TM if, for
every input string $x \in \Sigma^*$, the TM eventually halts and
produces the output $g(x)$. The function $g$ is said to be
computable in space bounded by $f(n)$ if, for all inputs $x =
x_1\cdots x_n$, the work tape head remains within the first $f(n)$
cells of the work tape throughout the computation. For decision
problems, where the answer for each $x_1\cdots x_n$ is 0 or 1, one
often disregards the output tape entirely and takes the first cell
of the work tape, say, to hold the answer whenever the computation
halts.

Of particular interest to this paper are functions computable
within space bounds of the form $f(n)=O(\log n)$, which is
significantly smaller than the space required to hold the entire
input. Functions that are computable within such a bound are
called {\em log-space computable} functions. Note that a function
$g$ computable in log-space need not have correspondingly short
outputs $y_1\cdots y_m = g(x_1\cdots x_n)$, because a log-space
bounded machine may still run for $O(\poly (n))$ steps
\cite{papadim,AB} (albeit in cramped space conditions) before
halting, and so we can potentially have $m=O(\poly (n))$ as well.

A general notion and study of space-bounded {\em quantum}
computation was given in \cite{wat1,wat2}. Here we will adopt a
simplified version of the model along the lines presented in
\cite{wat3}. Although one may seek to directly generalise the
classical definition above, using a fully quantum notion of a
quantum Turing machine of the kind described in \cite{bv}, it
appears to be much easier to work instead with a classical-quantum
hybrid TM model in which some aspects of the machine are required
to always remain classical.

We start with a {\em classical} TM as above, and add an additional
tape (called the {\em quantum work tape}) with the following
features:
\begin{enumerate}
\item[(i)] the cells of the quantum work tape are qubits, each of
  which is initialised to $\ket{0}$ at the start of the computation,
  and
\item[(ii)]
  there are {\em two} heads, with classical positions, scanning the
  quantum work tape.
\end{enumerate}
We will also eliminate the output tape as our focus will be on
decision problems, although one could make further modifications
to the model to allow for a classical output tape as well.

The operation of the machine is defined by a finite set of
instructions similar to eq.~(\ref{clinstr}), but suitably modified
so that quantum gates may be applied to the qubits on the quantum
tape during the course of the computation. In particular, we may
assume that each internal state $q$ has associated to it a
two-qubit quantum gate $U_q$, chosen from a universal set (which
should include the identity operation for the programmer's
convenience). Then a basic transition rule for the machine has the
form:
\begin{equation}\label{qtm}
  (q,s_\text{in},s_\text{work}) \rightarrow
  (r,t_\text{work},d_\text{input},d_\text{work},d_{\text{q1}},d_\text{q2}).
\end{equation}
The meanings of such a transition is similar to the ones described
previously, except that $d_\text{q1}$ and $d_\text{q2}$ denote
movement instructions for the two quantum work tape heads. The
computation proceeds in a similar way as before---but with the
addition that if the two quantum work tape heads are scanning
different squares at the start of a given step, then the two-qubit
operation $U_q$ is applied to the two scanned qubits before the
tape heads move. In the event that the machine enters the state
$q_\text{halt}$, the leftmost quantum work tape qubit is measured
in the computational basis to determine the output of the
computation.

The quantum computation is said to occur within space bound $f(n)$
if, for every input $x_1\cdots x_n$ given on the input tape, the
work tape head and qubit tape heads remain within the first $f(n)$
cells of their respective tapes throughout the computation.
Because our output is a single bit, this provides a notion of a
decision problem being computed by an $f(n)$-space-bounded quantum
computation.

It is important to point out that a space $f(n)$ bounded quantum
computation may be alternatively thought of as just a circuit of
quantum gates applied to $f(n)$ qubit lines, all initialised to
$\ket{0}$, where the circuit is determined by a classical space
$f(n)$ bounded computation on the input $x_1\cdots x_n$, i.e.~a
``space-$f(n)$ generated'' family of quantum circuits. Indeed, in
our hybrid classical-quantum TM, the quantum work tape may be
replaced by a classical output tape just as in the fully classical
model, and instead of applying quantum operations, the
corresponding gate names and line numbers are sequentially written
on the output tape.

Note that apart from the final measurement, all our quantum
operations are required to be {\em unitary} gates. One may
entertain a more general scenario in which general non-unitary
quantum operations (completely positive trace preserving maps) on
two qubits are allowed, as well as a mechanism for measurements of
qubits to occur during the computation that can influence the
classical parts of the machine. In the ubiquitous scenario of
polynomial {\em time} and polynomial width quantum computations,
say in terms of polynomial-sized quantum circuits, this provides
no further generality because arbitrary quantum operations may
always be represented via unitary operations with the inclusion of
extra ancillary qubit lines. However, for space-bounded
computation this generalisation can be non-trivial. For example, a
log-space computation with general non-unitary gates could involve
polynomially many steps, and to simulate such a computation by a
unitary process in the most straightforward way would hence
require a polynomial number of ancillary qubit lines, which are
not available in log-space. The general notion of space-bounded
quantum computation with general non-unitary gates has been
considered in \cite{wat1,wat2}, but for the purpose of the present
paper we restrict ourselves to the model with unitary gates, which
we accordingly call space-bounded {\em unitary} quantum
computation.

\subsection*{Acknowledgments}
AM and RJ contributed comparably to this work as the first authors.

AM, RJ, and BK acknowledge support from the EC network
QICS which provided collaborative opportunities for this work. AM
acknowledges helpful discussions with D.~Gottesman, suggesting the
technique used in eq. (\ref{vave}), and with S.~Wehner. AM and BK
are grateful to H.J.~Briegel for his support. RJ is supported by
UK's EPSRC funded QIP IRC and the EC network QAP. JW is supported
by Canada's NSERC, the Canadian Institute for Advanced Research
(CIFAR), and the QuantumWorks Innovation Platform. BK acknowledges
support of the FWF (Elise Richter Program). The research at the
Perimeter Institute is supported by the Government of Canada
through Industry Canada and by Ontario-MRI.

\end{document}